\documentclass[12pt]{book}

\usepackage[dvips]{graphicx,color}
\usepackage{makeidx,tsukuba}

\makeauthorindex
\makeindex

\begin{document}

\BookTitle{\itshape The 28th International Cosmic Ray Conference}
\CopyRight{\copyright 2003 by Universal Academy Press, Inc.}
\pagenumbering{arabic}

\def\simleq{\; \raise0.3ex\hbox{$<$\kern-0.75em \raise-1.1ex\hbox{$\sim$}}\; }
\def\simgeq{\; \raise0.3ex\hbox{$>$\kern-0.75em \raise-1.1ex\hbox{$\sim$}}\; }

\chapter{
Correlations and Charge Composition of UHECR without Knowledge of
Galactic Magnetic Field}

\author{ Peter Tinyakov$^{1,3}$ and Igor Tkachev$^{2,3}$ \\
{\it (1) Institute of Theoretical Physics, University
  of Lausanne, CH-1015 Lausanne, Switzerland\\ 
(2) CERN Theory  Division, CH-1211 Geneva 23, Switzerland\\
(3) Institute for Nuclear Research, 60th October Anniversary prospect
  7a, 117312 Moscow, Russia
} }

\section*{Abstract}
We develop a formalism which allows to study correlations of charged
UHECR with potential sources without using any Galactic
Magnetic Field (GMF) model. The method is free of subjective chose of
parameters on which the significance of correlations depends
strongly. We show that correlations of the AGASA dataset with BL Lacs
(found previously after reconstruction of particle trajectories in a
specific GMF) are present intrinsically and can be detected
without reference to a particular model of magnetic field.

\section{Introduction}
Correlations between UHECR and BL Lacs [7] improve noticeably after
correction of UHECR arrival directions for deflections in GMF [8].
The GMF model and its parameters were chosen in Ref. [8] following the
literature where they were obtained by fitting to the observed Faraday
rotation measurements. Because of the uncertainties in the model of
magnetic field, it is clearly desirable to have an alternative
procedure capable of detecting charged correlations of UHECR without
referring to a particular GMF.

Any {\it chance coincidences} between cosmic rays and potential
sources should be distributed over the sky according to the local
density of sources and exposure of a cosmic ray experiment.  Any
significant deviations from this distribution gives independent
signature that the correlations are real and should reflect physical
effects. There are {\em a priori} reasons to expect such deviations
from ``uniformity'' for real signal. The extra-galactic magnetic fields
are unlikely to be small in all directions [1]. If primary particles are
protons, one may expect good correlation with sources in some areas of
the sky and no correlations in the other. Poor knowledge of the
Galactic magnetic field may have similar effect: the directions of
cosmic rays before they enter the Galactic magnetic field may be
obtained correctly only in the regions where actual GMF is described
well by the model used.  Thus, one may expect that correlating rays
will not cover the acceptance region uniformly, but will form {\it
  spots} where the rate of correlations is high, while in other areas
the number of correlating rays will not exceed the random background.
Such spots in distribution of correlating rays were indeed found [9].
But are they due to deviations of GMF from a model in certain regions,
or the reason is different? The method which is insensitive to the GMF
model may give answer to such questions.

Why should such a procedure exist at all? Deflections by GMF form a
regular vector field; they are expected to point in close directions
in relatively small regions of the sky.  On the contrary, if
extra-galactic fields dominate the deflections, or if the association
between cosmic rays and ``sources'' is due to a mere coincidence, the
vectors of deflections would form a random field. In this talk we
develop the statistical test to distinguish between these two
situations, and confront it with the AGASA data, Ref. [3].

\section{Formulation of the procedure}

The procedure which we propose consists in the following steps: (1)
Choose reasonably small region on the sky (the domain where the
deflections are expected to point roughly in the same direction). (2)
Within this domain, identify pairs source -- cosmic ray by choosing
rays which have nearest source not further than maximum expected
deflection angle; this defines a set of directions of deflections
corresponding to the data. (3) Perform Kolmogorov-Smirnov (KS) test to
compare these directions with the ones obtained for the same domain,
the same set of sources, but randomly generated cosmic rays. The test
which is reparametrization-invariant on the circle should be used.

Consider this procedure in more detail in the case of AGASA cosmic
rays with energy $E>4\times10^{19}$~eV [3] and the set of confirmed BL
Lacs with $\mbox{mag}<18$ (the same set as in Refs. [8,9]):

(1) First, we have to divide the sky in regions. A natural choice is
the regions introduced previously in Ref. [9]. There, the part of the
sky overlapping with AGASA acceptance region was divided into 4 equal
area domains corresponding to north/south and inner/outer
Galaxy. These regions are labelled as follows:
\begin{center}
\begin{tabular}{lll}
 I:   & $0^{\circ}  < l  < 120^{\circ}$, & $b>0$\\
 II:  & $120^{\circ}< l  < 240^{\circ}$, & $b>0$\\ 
 III: & $120^{\circ}< l  < 240^{\circ}$, & $b<0$\\
 IV:  & $0^{\circ}  < l  < 120^{\circ}$, & $b<0$ 
\end{tabular}
\end{center}
For a particular GMF model adopted in Ref. [8], significant
correlations between cosmic rays and BL Lacs were found in regions I
and III, while no signal was observed in regions II and IV, see [9].
Note that in all existing models the deflections point roughly in the
same directions throughout region II or III (outer Galaxy), but field
of deflections is complicated in regions I and IV (inner Galaxy).

(2) To proceed, we have to find a set of pairs source -- cosmic ray
which define the deflections $\delta {\bf n}_i \equiv \left({\bf
  n}_i^{\rm cr} - {\bf n}_i^{\rm src}\right)/\mid {\bf n}_i^{\rm cr}
- {\bf n}_i^{\rm src}\mid$, where ${\bf n}_i^{\rm cr}$ and ${\bf
  n}_i^{\rm src}$ are unit vectors pointing in the direction of i-th 
cosmic ray and its candidate source, respectively.  We
identify such pairs as follows. For a typical energy of cosmic ray $\sim
4\times 10^{19}$~eV and typical magnitude of GMF, the deflections are
of order $3-8$ degrees. So, we look for candidate sources within
$10^{\circ}$ of the rays (we have verified that the results change
insignificantly when this parameter is increased to $20^{\circ}$). We
select the closest source if there are many candidates, and reject the
ray if there are none within specified region. Thus obtained
deflections $\delta {\bf n}_i$ are projected onto a fixed direction
${\bf e}_l$ of constant Galactic latitude $b$. This defines the set
of angles, $\alpha_i \equiv \arccos \left(\delta {\bf n}_i \cdot {\bf
  e}_l\right)$, which are expected to be distributed roughly uniformly
from 0 to $2\pi$ if cosmic rays and sources are uncorrelated, or if
the deflections are due to a random field.

(3) Finally, we perform a KS test to compare this distribution to the
one obtained for a large number of randomly generated cosmic rays.
Since both distributions are defined on a circle, we use a cyclic
version of the KS test [4]. This eliminates the dependence on the
fixed direction ${\bf e}_l$ the deflections were projected onto.

\section{Results}

In the region III, we find that deflections are aligned: their
directions vary within $60^{\circ} \simleq \alpha \simleq 90^{\circ}$.
Corresponding cumulative distribution is shown in Fig. 1, where solid
and dotted curves represent the real data and Monte-Carlo simulation
with random cosmic rays, respectively. These two distributions are
incompatible at the significance level of $P = 4 \cdot 10^{-5}$.

One should be careful applying correlation analysis to highest energy
CR data. Namely, the AGASA dataset is autocorrelated on the scale of
$2.5^\circ$ [2,3,6,5] and this may cause spurious correlation effects.
Within present procedure it is meaningless to incorporate UHECR
clustering into Monte-Carlo generator (as it was done e.g. in Ref [7])
since the expected distribution will be uniform anyway. But we can
eliminate any effects related to clustering by replacing each cluster
in the real data (regardless of its multiplicity) by a single cosmic
ray located at the cluster center. Real significance can be only
higher compared to the one found with clusters being removed. We
carried out this procedure of cluster reduction (eliminating three
doublets in region III) and found $P_{\tiny \rm no \, clusters} = 1.3
\cdot 10^{-4}$ in cyclic KS test.

In regions I and II we also find that the data deviate from
uncorrelated distribution, with KS significance $\approx 10^{-2}$ in
each region.  In region IV we see no signal (note however that this
region has the smallest number of AGASA CRs).

\section{Conclusions}

In region III the AGASA data are incompatible with uncorrelated
distribution; the significance is $4 \cdot 10^{-5} < P < 10^{-4}$.
Overall deflection angle is consistent with expectation derived in GMF
models with small vertical component of the magnetic field. In region
II the deflections are also expected to be aligned; smaller signal in
this region may be due either to a fluctuation, or to the effect of
extra-galactic magnetic fields. 

\begin{figure}[t]
  \begin{center}
    \includegraphics[height=15pc]{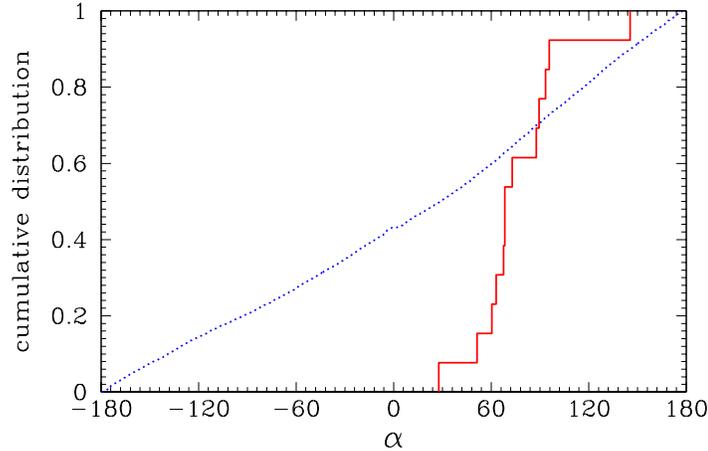}
  \end{center}
  \vspace{-0.5pc}
  \caption{Cumulative distribution of angles $\alpha_i$ 
    between cosmic ray deflections in region~III and direction ${\bf
      e}_l$ of constant Galactic latitude $b$. Dotted line -
    Monte-Carlo simulation of uncorrelated rays, solid line - real
    data.}
\end{figure}

\section{References}
\vspace{\baselineskip}
\re
1. Berezinsky~V., Gazizov~A.~Z., Grigorieva~S.~I.,
astro-ph/0210095
\re
2. Hayashida~N. et al. 1996,
Phys.\ Rev.\ Lett.\  {77}, 1000
\re
3.\ Hayashida~N. et al. 1999,
Astrophys.\ J.\ {522}, 225 and 
astro-ph/0008102
\re
4.\ Press~W.~H. et al. 1992, {Numerical Recipes,} chapter 14 
(Cambridge\\ University Press)
\re
5. Takeda~M. et. al. 2001, Proc. of 27th ICRC, p. 341
\re
6.\ Tinyakov P.~G., Tkachev I.~I. 2001,
JETP Lett. 74, 1 
\re
7.\ Tinyakov~P.~G., Tkachev~I.~I. 2001,
JETP Lett.\  {74}, 445
\re
8.\ Tinyakov~P.~G., Tkachev~I.~I. 2002,
Astropart.\ Phys.\  {18}, 165
\re
9.\ Tinyakov~P.~G., Tkachev~I.~I. 2002, 
International Workshop on Extremely\\ High Energy Cosmic Rays, Wako, 
Japan, 5-6 Nov 2002,
hep-ph/0212223

\end{document}